\input harvmac
\input epsf

\Title{\vbox{\baselineskip12pt\hbox{}
\hbox{FU Berlin preprint}}}
{\vbox{\centerline{Universality Class of}
\vskip2pt\centerline{Confining Strings}}}
\centerline{M. Cristina Diamantini\footnote{$^*$}{Supported by
an A. v. Humboldt fellowship. On leave of absence from I.N.F.N. and
University of Perugia; e-mail: diamanti@einstein.physik.fu-berlin.de}, 
\ Hagen Kleinert
\footnote{$^{**}$}{e-mail: kleinert@physik.fu-berlin.de ~~
http://www.physik.fu-berlin/\~{}kleinert}} 
\centerline{and C. A. Trugenberger
\footnote{$^{***}$}{
e-mail: cat@kalymnos.unige.ch}}
\centerline{Institut f\"ur Theoretische Physik, Freie Universit\"at Berlin}
\centerline{Arnimalle 14, D-14195 Berlin, Germany}

\vskip .3in
\noindent
A recently proposed model of confining strings has a non-local
world-sheet action induced by a space-time Kalb-Ramond tensor field.
Here we show that, in the large-$D$ approximation, 
an infinite set of ghost- and
tachyon-free truncations of the derivative expansion of this action
all lead to $c=1$ models. Their infrared limit describes smooth
strings with world-sheets of Hausdorff dimension $D_H=2$ and
long-range orientational order, as
expected for QCD strings.

\Date{March 1999}

\lref\polcha{For a review see e.g. :
J. Polchinski, ``{\it Strings and QCD}",
contribution in Symposium on Black Holes, Wormholes
Membranes and Superstrings, H.A.R.C., Houston (1992); hep-th/9210045.}

\lref\polchbook{For a review see e.g.: 
J. Polchinski, ``{\it String 
Theory}", Cambridge University Press, Cambridge (1998).} 

\lref\polbook{For a review see e.g.: 
A. M. Polyakov, ``{\it Gauge Fields and Strings}", Harwood Academic
Publishers, Chur (1987).}

\lref\polrev{A. M. Polyakov, {\it Physica Scripta} {\bf T15} (1987) 191.}

\lref\rigid{A. M. Polyakov, {\it Nucl. Phys.} {\bf B268} (1986) 406;
H. Kleinert, {\it Phys. Lett.} {\bf B174} (1986) 335.}

\lref\polb{A. M. Polyakov, {\it Nucl. Phys.} {\bf B486} (1997) 23.}

\lref\polc{A. M. Polyakov, {\it Physica Scripta} {\bf T15} (1987) 191.}

\lref\pold{A. M. Polyakov, {\it Nucl. Phys. Proc. Supp.} {\bf 68}
(1998) 1 (hep-th/9711002), {\it The Wall of the Cave}, hep-th/9809057.}

\lref\kb{H. Kleinert and A. Chervyakov, {\it Phys. Lett. } {\bf B381}
(1996) 286.}

\lref\alvarez{O. Alvarez, {\it Phys. Rev. } {\bf D24} (1981) 440.}

\lref\braaten{E. Braaten, R. D. Pisarski and S. M. Tze, {\it Phys.
Rev. Lett.} {\bf  58} (1987) 93.}

\lref\larged{H. Kleinert, {\it Phys. Rev. Lett.} {\bf 58} (1987) 1915;
P. Olesen and S. K. Yang, {\it Nucl. Phys.} {\bf B283} (1987) 73.}

\lref\temprigid{H. Kleinert, {\it Phys. Lett.} {\bf B189} (1987)
187, {\it Phys. Rev} {\bf D40} (1989) 473.}

\lref\david{F. David and E. Guitter {\it Nucl. Phys.} {\bf B295}
(1988) 332, {Europhys. Lett. } {\bf 3} (1987) 1169.}

\lref\bz{E. Braaten and C. K. Zachos, {\it Phys. Rev. } {\bf D35} (1987) 1512.}

\lref\qt{F. Quevedo and C. A. Trugenberger, {\it Nucl. Phys.} {\bf B501}
(1997) 143.}

\lref\dqt{M. C. Diamantini, F. Quevedo and C. A. Trugenberger, {\it Phys. Lett.}
{\bf B396} (1997) 115;
D. Antonov, {\it Phys. Lett.} {\bf B427} (1998) 274, {\bf B428}
(1998) 346.}

\lref\dt{M. C. Diamantini and C. A. Trugenberger, {\it Phys. Lett.}
{\bf B421} (1998) 196; M. C. Diamantini and C. A. Trugenberger, {\it Nucl.
Phys.} {\bf B531} (1998) 151.}

\lref\kr{V. I.
Ogievetsky and V. I. Polubarinov, {\it Sov. J. Nucl. Phys.} {\bf 4}
(1967) 156; M. Kalb
and P. Ramond, {\it Phys. Rev.} {\bf D9} (1974) 2273.}

\lref\ahmodel{ M. I. Polikarpov, U.-J. Wiese
and M. A. Zubkov, {\it Phys. Lett.} {\bf B309} (1993); K. Lee, {\it Phys.
Rev.} {\bf D48} (1993) 2493;
P.Orland, {\it Nucl. Phys.} {\bf B428} (1994) 221.}

\lref\gr{I. Gradstheyn and I. M. Ryzhik, ``{\it Table of Integrals, Series
and Products}", Academic Press, Boston (1980).}

\lref\wi{E. Witten, {\it Phys. Lett. } {\bf B86} (1979) 283.}

\lref\ps{J. Polchinski and A. Strominger, {\it Phys. Rev. Lett.} {\bf 67}
(1991) 1681.}

\lref\pz{J. Polchinski and Z. Yang, {\it Phys. Rev.} {\bf D46} (1992) 3667.}

\lref\cara{J. C. Cardy and E. Rabinovici, {\it Nucl. Phys.} 
{\bf B205} (1986) 1.} 

\lref\dfj{B. Durhuus, J. Fr\"ohlich and T. Jonsson, {\it Nucl. Phys.}
{\bf B240} (1984) 453; B. Durhuus and T. Jonsson, {\it Phys. Lett.}
{\bf 180B} (1986) 385.}

\lref\zinn{For a review see e.g.: J. Zinn-Justin, ``{\it Quantum Field
Theory and Critical Phenomena}", Oxford University Press, Oxford (1996).}

\lref\aa{D. Antonov, {\it Phys. Lett.} {\bf B427} (1998) 274, {\bf B428}
(1998) 346.}

\lref\ab{D. Antonov and D. Ebert, "{\it Dual Formulation and Confining
Properties of the SU(2)-Gluodynamics}", hep-th/9902177.}

\lref\klqed{H. Kleinert, {\it Phys. Lett.} {\bf B246} (1990) 127,
{\it Int. J. Theor. Phys.} {\bf A7} (1992) 4693, {\it Phys. Lett.}
{\bf B293} (1992) 168.}

\lref\zigzag{I. Kogan and O. Solovev, {\it Phys. Lett.} {\bf B442} (1998)
136; E. Alvarez and C. Gomez, {\it String Representation of Wilson Loops}, 
hep-th/9806075, {\it Non-critical Confining Strings and the Renormalization
Group}, hep-th/9902012; 
I Kogan, {\it On Zigzag Invariant Strings}, hep-th/9901131.}

\lref\mavro{J. Ellis and N. Mavromatos, {\it Confinement in Gauge Theories
from the Condensation of World-Sheet Defects in Liouville Strings}, 
hep-th/9808172; P. Horava, {\it On QCD String Theory and AdS Dynamics}, 
hep-th/9811028.}

\lref\hyper{M. C. Diamantini, H. Kleinert and C. A. Trugenberger, {\it Phys.
Rev. Lett. } {\bf 82} (1999) 267.}

\lref\gn{See e.g.: D. Gross,
{\it Application of the Renormalization
Group to High-Energy Physics}, in ``Methods in Field Theory", R. Balian
and J. Zinn-Justin eds., North-Holland \& World Scientific, Singapore (1981).}

\lref\numerical{M. N. Chernodub, M. I. Polikarpov, A. I. Veselov and
M. A. Zubkov, {\it Phys. Lett. } {\bf B432} (1998) 182.}

\lref\amit{S. Elitzur, A. Giveon, E. Rabinovici, A. Schwimmer and 
G. Veneziano, {\it Nucl. Phys.} {\bf B435} (1995) 147.}

\lref\cardybook{For a review see: J. Cardy, {\it Scaling and 
Renormalization in Statistical
Physics}, Cambridge University Press, Cambridge (1996).}

\lref\uniterm{H. W. Bl\"ote, J. Cardy and M. Nightingale, {\it Phys.
Rev. Lett.} {\bf 56} (1986) 742; I. Affleck, {\it Phys. Rev. Lett.}
{\bf 56} (1986) 746.}

\lref\pole{A. M. Polyakov, {\it JETP Lett.} {\bf 12} (1970) 381.}

\lref\lterm{M. L\"uscher, K. Symanzik and P. Weisz, {\it Nucl. Phys.}
{\bf B173} (1980) 365; M. L\"uscher, {\it Nucl. Phys. } {\bf B180}
(1981) 317.}

\lref\aes{D. Antonov, D. Ebert and Y. Simonov, {\it Mod. Phys. Lett.}
{\bf A11} (1996) 1905.}

\lref\caselle{M. Caselle, R. Fiore, F. Gliozzi, M. Hasenbusch
and P. Provero, {\it Nucl. Phys. } {\bf B486} (1997) 245.}

\lref\pa{R. Pisarski and O. Alvarez, {\it Phys. Rev.} {\bf D26}
(1982) 3735.}

\lref\nesterenko{V. V. Nesterenko and N. R. Shvetz, {\it Z. Phys.}
{\bf C55} (1992) 265.}

\newsec{Introduction}
Nonwithstanding the large amount of evidence \polcha \  suggesting 
the possibility of a 
string description of quark confinement, a consistent model of non-critical
strings has yet to be found. The simplest possibility, provided by the
Nambu-Goto string can be consistently quantized
only in $D=26$ or $D\le 1$ dimensions due to the conformal anomaly 
\polchbook .  Large world-sheets in 
Euclidean space crumple and the model is
inappropriate to describe the expected smooth strings dual to QCD \polc .
In an attempt to cure this problem
a naively marginal term proportional to the
square of the extrinsic curvature was added to the action \rigid .
The resulting rigid string is, however, different from the Nambu-Goto string
only in the ultraviolet region, since the new term turned out to be
infrared irrelevant. Thus the rigidity did not really help in
preventing crumpling \david .

Recent progress in this field is based on two types of actions. A first
model of confining strings \refs{\polb, \qt} \ is based on an induced
string action which can be explicitly derived for compact QED \refs{\polb,
\dqt, \klqed } \ and for Abelian-projected $SU(2)$ \ab . In its non-local
formulation, the model was independently proposed in \kb . A second 
proposal \pold , analyzed further in \zigzag , is based on a string
action in a five-dimensional curved space-time with the quarks living
on a four-dimensional horizon. Both proposals, whose interrelation has 
been investigated in \mavro , enjoy the necessary
zigzag invariance of QCD strings. 

In its world-sheet formulation, the induced string possesses a 
non-local action with negative stiffness \refs{\kb, \dqt} \ just as
the world-sheets of magnetic strings of the
Abelian Higgs model in the London 
limit of infinite Higgs mass \refs{\klqed, \ahmodel} .
Such an action may be brought to a quasi-local form via
a derivative expansion of the interaction between the surface
elements. For a conventional renormalization group
study of the geometric properties of the
fluctuating world-sheets we 
a truncate this derivative expansion. This
makes the model non-unitary, but in
a spurious way. In the truncated action the
stiffness is negative, so that  
a stable truncation 
must include at least a sixth-order
term in the derivatives. In \hyper \ we have shown that 
this term has the desired properties of solving  
the infrared problem of Nambu-Goto and rigid strings
in the large-$D$ approximation. While perturbatively irrelevant, it
becomes relevant in the large-$D$ approximation, 
a phenomenon familiar from 3$D$
Gross-Neveu model \gn . It suppresses crumpling and the model has
an infrared-stable fixed point corresponding to a tensionless smooth string
whose world-sheet has 
 Hausdorff dimension $D_H=2$. The corresponding
long-range orientational order is caused by a frustrated antiferromagnetic
interaction between normals, a mechanism first recognized in \dt \ and
confirmed by recent numerical simulations \numerical .

The purpose of this paper is to
determine the universality class of confining
strings determined by the finite-size scaling of the Euclidean 
effective action of the model \hyper \ on a cylinder of (spatial)
circumference $R$ \cardybook . 
In the limit of large $R$ this takes the form
\eqn\mone{\lim_{\beta \to \infty} {S^{\rm eff}\over \beta} = 
{\cal T} R-{\pi c (D-2)\over 6R} + \dots \ ,}
for $(D-2)$ transverse degrees of freedom,
the universality class being encoded in the pure number $c$.
This suggests that the effective theory 
describing the infrared behaviour is
a conformal field theory (CFT) with central charge $c$ \uniterm .
In this case the number $c$ also fixes the L\"uscher term
\lterm \ in the quark-antiquark potential:
\eqn\zero{V(R) = {\cal T} R -{\pi c (D-2)\over 24R}+\dots \ .}

By interchanging $R$ in \mone \ with
the inverse temperature $\beta$ we obtain
immediately the low-temperature behaviour of the model. We shall give
an estimate of the deconfinement temperature as well as its range of 
validity.

Finally we shall generalize the results of \hyper \ to higher
truncations of the original non-local action and show that 
the universality class and the geometric properties of world-sheets
are largely independent of the level of the truncation, implying
the irrelevance of the truncation and the spurious non-unitarity
deriving from it altogether.

\newsec{Finite-size scaling}
The truncated world-sheet model of confining strings proposed in
\hyper \ is defined in Euclidean space by the action
\eqn\one{S= \int d^2{\xi} \sqrt{g}\ g^{ab} \ {\cal D}_a
x_{\mu }\left( t - s{\cal D}^2 +{1\over m^2} {\cal D}^4 \right) 
{\cal D}_b x_{\mu } \ , }
where $g$ and ${\cal D}_a$ represent, respectively, the determinant and the
covariant derivatives with respect to the 
induced metric $g_{ab} = \partial_a x_{\mu }\partial_b x_{\mu }$ on the
world-sheet ${\bf x}\left( \xi_0, \xi_1 \right)$. 
The first term represents a bare surface tension $2t$, while the second
accounts for rigidity with a stiffness parameter $s$ which is negative
when generated dynamically by a tensor field in four-dimensional
space-time \refs{\kb, \dqt} . The last term ensures the
stability of the model. Since it contains the square of the gradient
of the extrinsic curvature matrices it suppresses the formation of spikes
on the world-sheet. In the large-$D$ approximation it generates a string
tension proportional to the square mass $m^2$ which takes 
control of the fluctuations
where the orientational correlation die off. 
For $t,s, m \to 0$, one
reaches an infrared fixed-point describing tensionless smooth
strings with long-range orientational order \hyper .
While the model \one \ is
a toy version of the action induced by an antisymmetric tensor field, 
it is known \aes \ 
that QCD strings possess a curvature expansion of exactly this type.

In this paper we shall analyze the leading large-$D$ behaviour of
the effective action on a cylinder of (spatial) circumference $R$.
This is the extension to our model of the calculations \alvarez \ for
the Nambu-Goto string and \refs{\braaten, \larged} \ for the
rigid string. Contrary to these papers, however, we consider 
periodic boundary conditions as in \temprigid  \ in
order to avoid the problem of
a non-uniform saddle-point metric
pointed out in \refs{\alvarez, \braaten} .
In order to simplify analytic computations
we shall moreover 
equate the stiffness to its fixed-point value from
the beginning by setting $s=0$ in \one . 

The large-$D$ calculation requires the introduction of a
Lagrange multiplier matrix $\lambda ^{ab}$ enforcing the
constraint $g_{ab}=\partial_a x_{\mu }\partial _b x_{\mu }$.
The action \one \ is thus extended to
\eqn\two{S \to S+\int d^2\xi \sqrt{g} \ \lambda ^{ab}
\left( \partial_a x_{\mu}\partial_b x_{\mu } -g_{ab} \right) \ .}
The world-sheet is parametrized in a Gauss map as
$x_{\mu}(\xi) = \left( \xi_0, \xi_1, \phi_i (\xi)\right)$
with $i=2, \dots, D-1$. Here $-\beta/2 \le \xi_0 \le
\beta/2$ and $-R/2 \le \xi_1 \le R/2$ and $\phi _i
(\xi)$ describe the $D-2$ transverse fluctuations.
We look for a saddle-point with diagonal metric
$g_{ab}={\rm diag}\left( \rho_0, \rho_1 \right)$ and
Lagrange multiplier $\lambda^{ab} = \lambda
\ g^{ab}$. With this ansatz the extended action becomes
\eqn\three{\eqalign{S &= A_{\rm ext} \ \sqrt{\rho _0 \rho_1}
\left[ (t+\lambda ) {{\rho_0+\rho_1}\over {\rho_0 \rho_1}}
-2 \lambda \right] \cr
&+ \int d^2{\xi} \sqrt{g}\ g^{ab} \partial_a
\phi^i \left( t+\lambda +{1\over m^2} {\cal D}^4 \right) 
\partial_b \phi^i\ ,\cr }}
where $A_{\rm ext}=\beta R$ is the extrinsic, projected
area in coordinate space. By integrating over the
transverse fluctuations we get, in the limit $\beta \to \infty$,
an effective action
\eqn\four{\eqalign{S^{\rm eff} &= S_0 + S_1 \ ,\cr 
S_0 &= A_{\rm ext} \ \sqrt{\rho _0 \rho_1}
\left[ (t+\lambda ) {{\rho_0+\rho_1}\over {\rho_0 \rho_1}}
-2 \lambda \right] \ ,\cr
S_1 &= {D-2\over 4\pi}\ \beta \sqrt{\rho_0} \sum_{n=-\infty}^{+\infty}
\int_{-\infty}^{+\infty} 
d p_0 \ {\rm ln} \left[ p^2 \left( t+\lambda + 
{p^4 \over m^2}  \right) \right] \ ,\cr }}
where 
\eqn\five{p^2 \equiv p_0^2 + \omega _n^2 \ ,\qquad \qquad
\omega _n \equiv {2\pi \over R\sqrt{\rho_1}} n \ .}
By introducing the mass scale $\mu = \sqrt{m \sqrt{t
+\lambda}}$ we can rewrite the sums and integrals 
in the one-loop contribution $S_1$ as
\eqn\six{\eqalign{S_1 &= {D-2\over 4\pi}\ \beta \sqrt{\rho_0 } \left( S_1^0 + 
2 \ {\rm Re} \ S_1^{\mu } \right) \ ,\cr
S_1^0 &= \sum_{n=-\infty}^{+\infty} \int_{-\infty}^{+\infty} dx
\ m \ {\rm ln} \left( x^2 +{\omega_n^2 \over m^2} \right) \ ,\cr
S_1^{\mu } &= \sum_{n=-\infty}^{+\infty} \int_{-\infty}^{+\infty} dx
\ {\rm ln} \left( x^2 + \omega_n^2 +i\mu^2 \right) \ ,\cr}}
where Re denotes the real part.
We shall dispose of the ultraviolet divergences in
these quantities by analytic regularization. Defining first
the logarithm as   
${\rm ln}\ x = \left[ -(d/d\beta) \ x^{-\beta} \right]_{\beta =0}$,
and  using the analytic interpolation of the integral \gr 
\eqn\seven{\int_{-\infty}^{+\infty} dx \ {1\over \left( x^2+q^2 \right)^n} 
= q^{1-2n} \ {{\Gamma \left( {1\over 2} \right) \Gamma \left(
n-{1\over 2} \right)} \over \Gamma (n)} }
to any real $n$,   
leads to the following formula for the regularized integrals:
\eqn\eight{\int_{\rm reg} dx \ {\rm ln} 
\left( x^2+a^2 \right) = 2\pi a \ .}
The sums are then regularized by analytic continuation of
the formula $\sum_{n=1}^{\infty} n^{-z}$ $= \zeta (z)$
for the Riemann zeta function \gr . Using 
$\zeta (-1) = -1/12$ one obtains immediately
\eqn\nine{S_1^0 = - {2 \pi ^2 \over 3R\sqrt{\rho_1}} \ ,}
which leads to the well known results for
the Nambu-Goto \refs{\alvarez, \pa} \ and the
rigid \temprigid \ strings.

The computation of $S_0^{\mu }$ is a bit more involved.
First we represent the right-hand side of \seven \ by
the analytic continuation of the
following integral representation \gr \ of the gamma
function,
\eqn\ten{{1\over \left( x^2+q^2\right) ^s} = {1\over \Gamma (s)} 
\int_0^{\infty} dt \ t^{s-1} \ {\rm exp}\left[ - 
\left( x^2+q^2 \right) t\right] \ .}
We then substitute the sum in $S_1^{\mu }$ by an equivalent expression
by means of the duality relation \nesterenko
\eqn\eleven{\sum_{n=-\infty}^{+\infty} {\rm exp}\left( - n^2 t
\right) = \sqrt{\pi\over t} \ \sum_{n=-\infty}^{+\infty}
{\rm exp} \left( -{\pi ^2 n^2\over t} \right) \ .}
Separating out the $n=0$ term in the sum and using the
representation 
\eqn\twelve{K_{\nu } \left( 2\sqrt{\beta \gamma} \right) =
{1\over 2} 
\ \left( {\gamma \over \beta} \right)^{\nu \over 2} \int_0^{\infty}
dx \ x^{\nu -1} \ {\rm exp}
\left( -\gamma x -{\beta \over x} \right) }
of the modified Bessel function \gr \ in the
remainder we find
\eqn\thirteen{S_1^{\mu } = {\pi \mu ^2 R\sqrt{\rho _1}\over 4}
- \sum_{n=1}^{\infty} {4\sqrt{i\mu ^2}\over n} \ K_1 \left(
Rn\sqrt{i\mu ^2 \rho_1} \right) \ .}
Altogether, we obtain the
effective action on the cylinder:
\eqn\fourteen{\eqalign{S^{\rm eff} &= \beta R \ \sqrt{\rho _0 \rho_1}
\left[ (t+\lambda ) {{\rho_0 + \rho_1}\over {\rho_0 \rho_1}} -2\lambda
+{D-2\over 2} \ {\mu^2 \over 4}\right] \cr
&- {D-2\over 2} \ {\beta \sqrt{\rho_0}\over 2\pi} \left[ 
{2\pi^2\over 3R\sqrt{\rho_1}} + {\rm Re} \left[
\sum_{n=1}^{\infty} {8 \sqrt{i\mu ^2}\over n} \ K_1\left( nR\sqrt{
i\mu ^2 \rho_1}\right) \right] \right]\ .}}  
Being interested only in the large-$R$ behaviour, we may
neglect the exponentially small terms arising from the
Bessel functions and arrive at 
the relevant approximation to $S^{\rm eff}$
to be used in the remaining computation:
\eqn\fifteen{S^{\rm eff} = \beta R \ \sqrt{\rho _0 \rho_1}
\left[ (t+\lambda ) {{\rho_0 + \rho_1}\over {\rho_0 \rho_1}} -2\lambda
+{D-2\over 2} \ {\mu^2 \over 4}\right] 
- {D-2\over 2} {\pi \beta \over 3 R} \sqrt{\rho_0 \over \rho_1}\ .}

The factor $(D-2)$ in $S^{\rm eff}$ ensures that, for large $D$,
the fields $\lambda $, $\rho_0$ and $\rho_1$ are extremal and
satisfy thus the saddle-point (``gap'') equations
\eqn\sixteen{\eqalign{{{\rho_0+\rho_1}\over \rho_0\rho_1} -2+ {D-2\over 2}
{\mu^2 \over 8(t+\lambda)} &=0\ ,\cr
{{t+\lambda} \over 2} \left( {1\over \rho_1}-{1\over \rho_0} \right)
-\lambda + {D-2\over 2} {\mu^2 \over 8} - {D-2\over 2} {\pi \over
6R^2\rho_1} &=0\ ,\cr
{{t+\lambda} \over 2} \left( {1\over \rho_0}-{1\over \rho_1} \right)
-\lambda + {D-2\over 2} {\mu^2 \over 8} + {D-2\over 2} {\pi \over
6R^2\rho_1} &=0\ .\cr}}
Substituting the second of these equations in \fifteen \ we obtain the
simplified form of the effective action
\eqn\seventeen{S^{\rm eff} = \beta R \ {\cal T}
\sqrt{\rho_1 \over \rho_0} \ ,}
with ${\cal T} \equiv 2(t+\lambda)$ being the physical string tension.

The saddle-point equations are easily solved as follows. 
The sum of the last
two equations yields an equation for $\lambda $ alone,
\eqn\eighteen{\lambda - {D-2\over 2} \ {\mu ^2\over 8}=0\ .}
Using $\mu ^2 = m \sqrt{t+\lambda }$ this leads to the
following solution for the string tension ${\cal T} =
2(t+\lambda)$:
\eqn\nineteen{\eqalign{{\cal T} &= {a^2 \over 32} \ \left( {D-2\over 2}
\right) ^2 \ m^2\ ,\cr
a^2 &\equiv {1+128 \left( {2\over D-2} \right)^2 
\ {t\over m^2} + \sqrt{1+ 256 \left( {2\over D-2} \right)^2
\ {t\over m^2}}\over 2} \ ,\cr}}
reproducing the result of \hyper .
In a second step we subtract the second equation from the third, 
and multiply the result by $2 \rho_1/{\cal T}$, obtaining
\eqn\twenty{{\rho_1\over \rho_0} = 1- {\pi (D-2)\over 3{\cal T}R^2}\ .}
Expanding the square-root of this expression and multiplying by
$\beta R {\cal T}$ we obtain the final result 
\eqn\twentyone{{S^{\rm eff}\over \beta} = {\cal T}R -{\pi (D-2)
\over 6R} +\dots \ .}
Thus we conclude that confining strings are characterized
by $c=1$. Although they share the same value of $c$, confining strings
are clearly different $c=1$ theories than Nambu-Goto or rigid strings.
Indeed the former are smooth strings on any scale while the latter 
crumple and fill the ambient space, at least in the infrared region. 
Our result $c=1$ is in agreement with recent precision numerical
determinations \caselle \ of this constant.

\newsec{The deconfinement temperature}

By changing $R$ into $\beta $ and $\rho_0$ into $\rho_1$ in the
above formulas we obtain the behaviour
of the model \one \ at temperature $T=1/k_B\beta$. Having 
neglected the contribution of the Bessel functions 
in \fourteen , however, we can only study low temperatures, 
whith $\beta \mu \sqrt{\rho _0} > 1$. 
Using \seventeen \ and \twenty \ we get
\eqn\twentytwo{\left( {S^{\rm eff}\over R} \right) ^2 
= \beta ^2 {\cal T}^2 - {\pi (D-2) {\cal T}\over 3} \ .}
Raising the temperature, this quantity, representing the square mass
of the lowest state, crosses zero at an inverse temperature
\eqn\twentythree{\beta _{\rm dec} = \sqrt{\pi (D-2)\over 3{\cal T}} = 
{1\over m} \sqrt{128 \pi \over 
3(D-2) a^2}\ ,}
which specifies the deconfinement temperature of the model.
Note that this result coincides with the corresponding one for
Nambu-Goto \pa \ and rigid \temprigid \ strings
when expressed in terms of the string tension.

In order to establish the range of validity of this result we need to
know the value of $\sqrt{\rho_0}$. This is obtained
by substituting \twenty \ into the first of the saddle-point equations
\sixteen , yielding 
\eqn\twentyfour{\rho _0 = {{2a \left( 1- {\pi (D-2)\over
6{\cal T}\beta ^2} \right)}\over {2a-1}} \ .}
At the deconfinement temperature this becomes
\eqn\twentyfive{\left( {\rho_0}\right) _{\rm dec} = 
{a \over 2a-1}\ .}
The value \twentythree \ of the deconfinement temperature is
consistent with our approximation only if the equation
$\beta_{\rm dec} \mu \sqrt{{\rho_0}_{\rm dec}} > 1$ is satisifed.
Only then can we neglect the Bessel
functions down to the deconfinement transition. Using the
above value of $\beta _{\rm dec}$ and ${\rho_0}_{\rm dec}$ this
condition translates into 
\eqn\twentysix{a < \left( {8\pi \over 6}-{1\over 2} \right) 
\simeq 3.5 \ .}
Thus, formula \twentythree \ for the deconfinement temperature
is reliable in the region of small $(t/m^2)$ where 
$a\simeq 1$. Otherwise
there are sizable corrections from the sum over $n$ in \fourteen .

\newsec{Generalization to higher truncations}
Having established that the model \one \ describes smooth strings 
with $c=1$, the question arises as to how much these results depend on
the truncation of the original non-local action after the
${\cal D}^4$ term. To answer
this question let us consider instead of
\one \ an arbitrary truncation
\eqn\twentyseven{\eqalign{S|_n &= \int d^2\xi \sqrt{g} \ g^{ab} 
\ {\cal D}_a x_{\mu } \ V_n\left( {\cal D}^2 \right) 
\ {\cal D}_b x_{\mu } \ ,\cr
V_n\left( {\cal D}^2 \right) &= \left( \alpha _0+\lambda \right)
\Lambda ^2 +\sum_{k=1}^{2n} \ {\alpha _k\over \Lambda ^{2k-2}}
{\cal D}^{2k} \ .\cr }}
Here $\Lambda $ represents the fundamental mass scale in the model,
to be identified with the QCD mass scale, and we have already included
in the action the Lagrange 
multiplier $\lambda $ arising from \two \ (note that here we
have defined $\lambda $ as a dimensionless quantity). 
Since all expansion coefficients $\alpha _k$ are positive, the series
is alternating in momentum space,
with all terms with odd index $k$ being negative 
\refs{\kb, \dt} . Thus, stable
truncations must end with an even $k=2n$.

Following \hyper , the only condition we shall impose on the 
coefficients $\alpha _k$ is the absence of both tachyons and ghosts.
This requires that the Fourier-transform $V_n\left( p^2 \right)$ has
no zeros on the real $p^2$-axis. The polynomial $V_n \left( p^2 \right)$
has thus $n$ pairs of complex-conjugate zeros in the complex $p^2$-plane.

For simplicity of computation we shall set all coefficients with odd $k$
to zero, $\alpha _{2m+1} =0$ for $0\le m \le n-1$. This, however, is
no drastic restriction since, as we shall now demonstrate, this is
their value at the infrared-stable fixed point anyhow. With this
simplification all pairs of complex conjugate zeros of $V_n\left( p^2
\right)$ lie on the imaginary axis and we can represent $V_n
\left( p^2 \right)$ as
\eqn\twentyeight{{\Lambda ^{4n-2}\over \alpha _{2n}} \ V_n\left( p^2
\right) = \prod _{k=1}^n \left( p^4 + \gamma _k^2 \Lambda ^4 \right) \ ,}
with purely numerical coefficients $\gamma _k$. This 
expression substitutes $\left[ p^4 + m^2 (t+\lambda) \right]$
inside the logarithm in the one-loop contribution \four , which becomes
\eqn\twentynine{\eqalign{S_1 &= {D-2\over 4\pi} \ \beta \sqrt{\rho_0}
\left( S_1^0 + \sum_{k=1}^n \ 2 \ {\rm Re} \ S_1^k \right) \ ,\cr
S_1^k &= \sum_{l=-\infty}^{+\infty} \int_{-\infty}^{+\infty}
dx\ {\rm ln} \left( x^2 + \omega_l^2 + i\gamma _k \Lambda ^2 
\right) \ .\cr}}
Using \thirteen \ and neglecting as before the Bessel functions o
for large $R$, we see that the only modification
to \fifteen , \sixteen  \ and \seventeen \ due to the higher-order
truncation is the substitution 
\eqn\thirty{\mu ^2 \to \sum_{k=1}^n \ \gamma_k \Lambda ^2 \ .}
The Lagrange multiplier $\lambda \Lambda ^2$ and the string tension
${\cal T} = 2\left( \alpha_0 +\lambda \right) \Lambda ^2$ are now
determined by the new saddle-point equation
\eqn\thirtyone{\lambda - {D-2\over 16} \ \sum_{k=1}^n \gamma_k =0 \ .}
The value $c=1$ for the universal term in \twentyone , however,
remains {\it unchanged}.

The new saddle-point equation for $\lambda $ is still polynomial,
although of higher-order. The requirement that this polynomial
``gap " equation has at least one solution on the real axis with
$\left( \alpha_0 +\lambda \right) \ge 0$ provides the condition
on the coefficients $\alpha _{2m}$, $0\le m \le n$, that defines
the universality class of confining strings at level $n$. 

Note that with all $\alpha _{2m+1}=0$ for $0\le m \le n-1$, no
normalization scale needs to be introduced to define the one-loop
term $S_1$. In other words, a scale introduced to properly
define the logarithm in \four \ would drop out at the end of the
computation since the result does not contain logarithms.
As a consequence, in a
renormalization analysis as in \hyper , there are no anomalous
dimensions and the infrared limit $\Lambda ^2 \to 0$
of vanishing string tension is characterized by $\beta \left(
\alpha_{2m} \right) =0$ for $0\le m\le n$. The point $\Lambda =0$
is thus again an infrared-stable fixed point characterized by
$\alpha _{2m+1} =0$ for $0\le m\le n-1$, and $n+1$ renormalization
group invariant numerical coefficients $\alpha _{2m}$, $0\le m \le n$, 
varying in a range where there exists a real solution to the
``gap" equation.

The geometric properties of world-sheets in the vicinity of this
point can be easily obtained by decomposing
\eqn\thirtytwo{{1\over {V_n\left( p^2 \right) }} = {\Lambda ^2 \over
\alpha _{2n}} \ \sum_{k=1}^n 
\ {\eta _k \over {p^4+\gamma_k^2 \Lambda ^2}}\ .}
This decomposition is always possible since it is determined by
a linear system of $n$ equations for the $n$ numerical coefficents
$\eta_k$. At this point we can simply apply to each term in the
above decomposition the discussion
of \hyper \ and conclude that 
the infrared point of vanishing tension is characterized by long-range
orientational order and Hausdorff dimension $D_H=2$ of world-sheets. 

We have thus shown that $c$ and the smooth geometric properties
are independent of an infinite set of truncations, provided 
that a solution for the polynomial ``gap" equation exists. These
properties are presumably common to a large class of non-local
world-sheet interactions.

\listrefs
\end